\documentclass[preprint,showpacs,preprintnumbers,amsmath,amssymb,aps]{revtex4}
\usepackage{hyperref}
\usepackage{graphics}
\usepackage{graphicx}
\usepackage{color}
\usepackage{amsmath}
\usepackage{amssymb}
\usepackage{bm}
\usepackage{float}
\usepackage{epsfig}
\usepackage{epstopdf}
\usepackage{color,graphics}

\newcommand\be{\begin{equation}}
\newcommand\e{\end{equation}}
\newcommand\ba{\begin{eqnarray}}
\newcommand\ay{\end{eqnarray}}

\begin{document}

\title{Linear and nonlinear ion-acoustic waves in non-relativistic quantum
plasmas with arbitrary degeneracy}
\author{Fernando Haas}
\affiliation{Physics Institute, Federal University of Rio Grande do Sul, CEP 91501-970,
Av. Bento Gon\c{c}alves 9500, Porto Alegre, RS, Brazil}
\author{Shahzad Mahmood}
\affiliation{Physics Institute, Federal University of Rio Grande do Sul, CEP 91501-970,
Av. Bento Gon\c{c}alves 9500, Porto Alegre, RS, Brazil}
\affiliation{Theoretical Physics Division (TPD), PINSTECH, P. O. Nilore Islamabad 44000,
Pakistan}

\begin{abstract}
Linear and nonlinear ion-acoustic waves are studied in a fluid model for
non-relativistic, unmagnetized quantum plasma with electrons with an
arbitrary degeneracy degree. The equation of state for electrons follows
from a local Fermi-Dirac distribution function and apply equally well both
to fully degenerate or classical, non-degenerate limits. Ions are assumed to
be cold. Quantum diffraction effects through the Bohm potential are also
taken into account. A general coupling parameter valid for dilute and dense
plasmas is proposed. The linear dispersion relation of the ion-acoustic
waves is obtained and the ion-acoustic speed is discussed for the limiting
cases of extremely dense or dilute systems. In the long wavelength limit the
results agree with quantum kinetic theory. Using the reductive perturbation
method, the appropriate Korteweg-de Vries equation for weakly nonlinear
solutions is obtained and the corresponding soliton propagation is analyzed.
It is found that soliton hump and dip structures are formed depending on the
value of the quantum parameter for the degenerate electrons, which affect
the phase velocities in the dispersive medium.
\end{abstract}

\pacs{52.35.Fp, 52.35.Sb, 67.10.Db}
\maketitle

\section{Introduction}

The study of degenerate plasma is important due to its applications e.g. to
strong laser produced plasmas \cite{r2}, high density astrophysical plasmas
such as in white dwarfs or neutron stars \cite{r5}, or large density
electronic devices (as in the drain region of $n^{+}nn^{+}$ diodes \cite{r1}%
). In plasmas, the quantum effects are more relevant for electrons rather
than ions because of their lower mass. The quantum nature of the charge
carriers manifests with the inclusion of both Pauli exclusion principle for
fermions and Heisenberg uncertainty principle due to the wave like character
of the particles. Accordingly, electrons obey the Fermi-Dirac statistics and
their equation of state is determined using the Fermi-Dirac distribution. On
the other hand the quantum diffraction effects are usually modeled by means
of quantum recoil terms in kinetic theo\-ry or the Bohm potential in fluid
theory, besides higher order gradient corrections \cite{r8, mel}.

Accordingly, the wave propagation in a degenerate plasma can be studied
using at least two main approaches i.e., kinetic and hydrodynamic models. In
kinetic theory, the unperturbed electron distribution is frequently given by
a Fermi-Dirac function, while in hydrodynamics the momentum equation for
electrons is made consistent with the equation of state of a degenerate
electron Fermi gas \cite{r8, mel}. In fluid models, the ion-sound wave
pro\-pa\-ga\-tion in plasmas with degenerate electrons has been investigated
by a number of authors \cite{r12,r13,r14,r15,r16,r17,r18}, using the
equation of state for a cold (fully degenerate) Fermi electron gas, with a
negligible thermodynamic temperature. The energy distribution of a
degenerate electron gas des\-cri\-bed by the Fermi-Dirac distribution is
characterized by independent parameters, one of which is the chemical
potential, while the other is the thermodynamic tem\-pe\-ra\-tu\-re. On the
other hand, the energy spread for the classical ideal electron gas obeying
Maxwell-Boltzmann distribution is uniquely determined by the thermodynamic
temperature. The equation of state for the fully degenerate electron gas so
reduces to an one-parameter problem i.e., the chemical potential. Therefore,
it is of interest to study the linear and specially nonlinear wave
propagation in the intermediate regime, depending on the competition between
the two parameters i.e., chemical potential and thermal temperature \cite%
{r11}, including quantum diffraction.

Our treatment is specially relevant to borderline systems with $T \approx
T_F $, which are neither strongly degenerate nor sufficiently well described
by classical statistics, where $T$ and $T_F$ are resp. the electron
thermodynamic and Fermi temperatures. A striking example is provided by
inertial confinement fusion plasmas \cite{solid}, with particle densities
ranging from $10^{30} m^{-3}$ to $10^{32} m^{-3}$, and thermodynamic
temperatures above $10^7 K$. During laser irradiation of the solid target,
quantum statistical effects tend to be more relevant immediately after
compression, before the heating phase. Moreover, laboratory simulation of
astrophysical scenarios involving dense plasmas better fit the intermediate
quantum-classical regime \cite{gregori}. For these reasons and potential
applications on e.g. ultra-small semiconductor devices operating in a mixed
dense-dilute regime \cite{r1}, it is desirable to have a general
ma\-cros\-co\-pic model covering both classical and quantum statistics,
besides quantum diffraction.

Previously, Maafa \cite{r19} studied the ion-acoustic and Langmuir waves in
a plasma with arbitrary de\-ge\-ne\-ra\-cy of electrons using classical
kinetic theory, linearizing the Vlasov-Poisson system around a Fermi-Dirac
equilibrium. Using quantum kinetic theory, Melrose and Mushtaq derived the
electron-ion plasma low-frequency longitudinal response including quantum
recoil, first for dilute (Maxwell-Boltzmann equilibrium) plasmas \cite{r22}
and then \cite{mm} for general degeneracy, in a Fermi-Dirac
e\-qui\-li\-brium. These works were restricted to linear waves only.
Eliasson and Shukla \cite{r23} derived nonlinear quantum electron fluid
equations by taking the moments of the evolution equation for the Wigner
function in terms of a local Fermi-Dirac equilibrium with an arbitrary
thermodynamic temperature. In this model quantum diffraction manifest in
terms of the Bohm potential. The high (classical) as well as the low
(degenerate) temperature limits of the obtained fluid equations were also
discussed in connection to Langmuir waves. Recently Dubinov \textit{et al.} 
\cite{r24} investigated the nonlinear theory of ion-acoustic waves in
isothermal plasmas with arbitrary degeneracy, but without including quantum
recoil. They presented the 
equation of state for ions and electrons by considering them as warm ($T
\neq 0$) Fermi gases. The nonlinear analysis was done using a Bernoulli
pseudo-potential approach. The ranges of the phase velocities of the
periodic ion-acoustic waves and the soliton velocity were investigated.
Ho\-we\-ver, for simplicity they ignored the quantum Bohm potential, which
increases the order of the resulting dynamical equations. Our central issue
here is to analyze the combined quantum statistical and quantum diffraction
effects on linear and nonlinear ion-acoustic structures in plasmas, in an
analytically simple (but hopefully not simplistic) approach.

The manuscript is organized in the following way. In Section II, the basic
set of hydrodynamic equations is proposed and the barotropic equation of
state defined, for a general Fermi-Dirac equilibrium. In Section III, the
linear dispersion relation for quantum ion-acoustic waves is derived,
following the fluid model. Comparison with known results from quantum
kinetic theory allows to determine a fitting parameter in the quantum force,
so that the macroscopic and microscopic approaches coincide in the long
wavelength limit. In Section IV, nonlinear wave structures are studied by
means of the reductive perturbation method and the associated Korteweg-de
Vries (KdV) equation. The associated quantum soliton solution is obtained.
Section V studies the possibility of forward and backward propagating
solitons in real systems. Finally, Section VI collect some conclusions.

\section{Model equations}

In order to study ion-acoustic waves in unmagnetized electron-ion plasmas
with arbitrary electron temperature, we use the set of dynamic equations
described as follows \cite{r8}.

The ion continuity and momentum equations are res\-pec\-ti\-ve\-ly given by 
\begin{eqnarray}
\frac{\partial n_{i}}{\partial t}+\frac{\partial }{\partial x}(n_{i}u_{i})
&=&0\,,  \label{e1} \\
\frac{\partial u_{i}}{\partial t}+u_{i}\frac{\partial u_{i}}{\partial x} &=&-%
\frac{e}{m_{i}}\frac{\partial \phi }{\partial x}\,.  \label{e2}
\end{eqnarray}%
The momentum equation for the inertialess quantum electron fluid is given by 
\begin{equation}
0=e\frac{\partial \phi }{\partial x}-\frac{1}{n_{e}}\frac{\partial p}{%
\partial x}+\frac{\alpha \,\hbar ^{2}}{6\,m_{e}}\frac{\partial }{\partial x}%
\left( \frac{1}{\sqrt{n_{e}}}\frac{\partial ^{2}}{\partial x^{2}}\sqrt{n_{e}}%
\right) .  \label{e3}
\end{equation}%
The Poisson equation is written as 
\begin{equation}
\frac{\partial ^{2}\phi }{\partial x^{2}}=\frac{e}{\varepsilon _{0}}%
(n_{e}-n_{i})\,,  \label{e4}
\end{equation}%
where $\phi $ is the electrostatic potential. The ion fluid density and
velocity are represented by $n_{i}$ and $u_{i}$ respectively, while $n_{e}$
is the electron fluid density. Also, $m_{e}$ and $m_{i}$ are the electron
and ion masses, $-e$ is the electronic charge, $\varepsilon _{0}$ the vacuum
permittivity and $\hbar $ the reduced Planck constant. In Eq. (\ref{e3}), $%
\alpha $ is a dimensionless constant factor to be determined and $p=p(n_{e})$
is the electron's fluid scalar pressure, to be specified by a barotropic
equation of state obtained in the continuation.

The last term proportional to $\hbar^2$ on the right hand side of the
momentum equation for electrons is the quantum force, which arises due to
the quantum Bohm potential, responsible for quantum diffraction or quantum
tunneling effects due to the wave like nature of the electrons. The
dimensionless quantity $\alpha$ will be selected in order to exactly fit the
kinetic theory linear dispersion relation in a three-dimensional Fermi-Dirac
equilibrium, as detailed in Section III. It is known that the qualitative
role of the Bohm potential is to provide extra dispersion. Ho\-we\-ver, the
precise numerical coefficient in its definition is a debatable subject
involving e.g. the dimensionality and the temperature \cite{Barker}. For
instance, for a local Maxwell-Boltzmann equilibrium, Gardner \cite{Gardner}
has found a factor $\alpha = 1$. Frequently, the factor $\alpha$ is set in
order to fit numerical results from kinetic theory \cite{Grubin}, which is
in the spirit of the present work. On the other hand, quantum effects on
ions are ignored in the model in view of their large mass. For simplicity,
ion temperature effects are also disregarded.

In order to derive the equation of state, consider a local quasi-equilibrium
Fermi-Dirac particle distribution function $f = f(\mathbf{v},\mathbf{r},t)$
for electrons \cite{r26}, given by 
\begin{equation}
f(\mathbf{v},\mathbf{r},t) = \frac{A}{1 + e^{\beta(\varepsilon - \mu)}} \,,
\label{e5}
\end{equation}%
where $\beta = 1/(\kappa_B T)$, $\varepsilon = m_{e}v^{2}/2, v = |\mathbf{v}%
| $ and $\mu$ is the chemical potential regarded as a function of position $%
\mathbf{r}$ and time $t$. Besides, $\kappa_B$ is the Boltzmann constant, $T$
is the (constant) thermodynamic electron's temperature and $\mathbf{v}$ is
the velocity. In addition, $A$ is chosen to ensure the normalization $%
\int\!f d^{3}v = n_e$, so that 
\begin{equation}
A = - \, \frac{n_e}{\mathrm{Li}_{3/2}(-e^{\beta \mu })}\left(\frac{\beta
m_{e}}{2\pi} \right) ^{3/2} = 2\left( \frac{m_{e}}{2\pi \hbar }\right) ^{3}
\,,  \label{e8}
\end{equation}
the last equality following from the Pauli principle (the factor two is due
to the electron's spin). Therefore, in the fluid description, $\mu$ and $A$
are supposed to be slowly varying functions of space and time. Equation (\ref%
{e8}) contains the poly-logarithm function $\mathrm{Li}_{\nu}(\eta)$ of
index $\nu$, which can be generically defined \cite{r27} by 
\begin{equation}
\mathrm{Li}_{\nu}(\eta) = \frac{1}{\Gamma(\nu)}\int_{0}^{\infty}\frac{s^{\nu
- 1}}{e^{s}/\eta-1} \,ds \,,
\end{equation}
where $\Gamma(\nu)$ is the Gamma function. We also observe that a
three-dimensional equilibrium is assumed, although for electrostatic wave
propagation only one spatial variable $x$ is needed in the model equations.

The scalar pressure follows from the standard definition for an equilibrium
with zero drift velocity, 
\begin{equation}
p = \frac{m_e}{3} \int \!f v^2 d^{3}v \,,
\end{equation}
yielding 
\begin{equation}  \label{e9}
p = \frac{n_{e}}{\beta}\frac{\mathrm{Li}_{5/2}(- e^{\beta\mu})}{\mathrm{Li}%
_{3/2}(- e^{\beta\mu})} \,.
\end{equation}

It is worth to consider some limiting cases of the barotropic equation of
state. From Eq. (\ref{e9}), in the dilute plasma limit case with a local
fugacity $e^{\beta\mu} \ll 1$ and using $\mathrm{Li}_{\nu }(-e^{\beta \mu })
\approx -e^{\beta \mu }$, one has 
\begin{equation}
p = n_{e}k_{B}T \,,  \label{e11}
\end{equation}%
which is the classical isothermal equation of state.

On the opposite, dense limit with a large local fugacity $e^{\beta \mu } \gg
1$, from $\mathrm{Li}_{\nu }(-e^{\beta \mu })\approx -\left( \beta \mu
\right) ^{\nu}/\Gamma (\nu +1)$ the result is 
\begin{equation}
p = \frac{2}{5}\,n_{0}\varepsilon_{F}\,\left(\frac{n_{e}}{n_{0}}\right)
^{5/3} \,,  \label{e12}
\end{equation}
which is the equation of state for a three-dimensional completely degenerate
Fermi gas, expressed in terms of the equilibrium number density $n_0$. In
Eq. (\ref{e12}), the electron's Fermi energy is $\varepsilon _{F} = \kappa_B
T_F = [\hbar^2/(2m_e)] \,(3\pi ^{2}n_{0})^{2/3}$, which is the same as the
e\-qui\-li\-brium chemical potential in the fully degenerate case. In
addition, $n_0$ is the equilibrium electron (and ion) number density.

The present treatment has similarities, as well as some different choices,
in comparison to Eliasson and Shukla work \cite{r23}. In this article, also
a local quasi-equilibrium Fermi-Dirac distribution function was employed.
Ho\-we\-ver, presently a non-constant chemical potential is admitted. In
addition, in Ref. \cite{r23} the focus was on situations involving
one-dimensional laser-plasma compression experiments, while here it is
assumed a three-dimensional isotropic equilibrium. Finally, the present work
deals with low-frequency (ion-acoustic) instead of high-frequency (Langmuir)
waves.

In passing, from Eq. (\ref{e8}) one deduce the useful relation 
\begin{equation}
n_{e} = n_{0}\,\frac{\mathrm{Li}_{3/2}(-e^{\beta\mu })}{\mathrm{Li}%
_{3/2}(-e^{\beta\mu_{0}})} \,,  \label{e18}
\end{equation}
where $\mu_0$ is the equilibrium chemical potential, satisfying 
\begin{equation}  \label{muu}
- \,\frac{n_0}{\mathrm{Li}_{3/2}(-e^{\beta \mu_0 })}\left(\frac{\beta m_{e}}{%
2\pi} \right) ^{3/2} = 2\left( \frac{m_{e}}{2\pi \hbar }\right) ^{3} \,.
\end{equation}

Using the equation of state (\ref{e9}), the chain rule and the property $d 
\mathrm{Li}_{\nu}(\eta)/d\eta = (1/\eta) \mathrm{Li}_{\nu-1}(\eta)$, the
momentum equation (\ref{e3}) for the inertialess electron fluid becomes 
\begin{eqnarray}
0 &=& e\frac{\partial\phi}{\partial x} - \frac{1}{\beta n_{e}}\frac{\mathrm{%
Li}_{3/2}(-e^{\beta\mu})}{\mathrm{Li}_{1/2}(-e^{\beta\mu})}\frac{\partial
n_{e}}{\partial x}  \notag \\
&+& \frac{\alpha\,\hbar^{2}}{6\,m_{e}}\frac{\partial}{\partial x}\left(\frac{%
1}{\sqrt{n_{e}}}\frac{\partial^{2}}{\partial x^{2}}\sqrt{n_{e}}\right) \,.
\label{e16}
\end{eqnarray}
Finally, using Eq. (\ref{e18}), we have the alternative form 
\begin{eqnarray}
0 &=& e\frac{\partial\phi}{\partial x} - \frac{1}{\beta n_{0}}\frac{\mathrm{%
Li}_{3/2}(-e^{\beta \mu _{0}})}{\mathrm{Li}_{1/2}(- e^{\beta\mu})}\frac{%
\partial n_{e}}{\partial x}  \notag \\
&+& \frac{\alpha\,\hbar^{2}}{6\,m_{e}}\frac{\partial}{\partial x} \left(%
\frac{1}{\sqrt{n_{e}}}\frac{\partial^{2}}{\partial x^{2}}\sqrt{n_{e}}
\right) \,,  \label{e19}
\end{eqnarray}
containing the minimal number of poly-logarithmic functions with a
non-constant argument.

It is worth noticing that the model does not include collisional damping,
which is reasonable if the average electrostatic potential per particle $%
\langle U \rangle$ is much smaller than the corresponding average kinetic
energy $\langle K \rangle$. For any degree of degeneracy, one can estimate $%
\langle U \rangle \approx e^{2}/(4 \pi \varepsilon_0 r_{S})$, where the
Wigner-Seitz ratio $r_S$ is defined by $(4 \pi r_{S}^3/3) \,n_0 = 1$. On the
other hand, from $\langle K \rangle = [m_e/(2n_e)]\int f v^2 d^{3}v$ and
evaluating on e\-qui\-li\-brium, one derive the general coupling parameter 
\begin{eqnarray}
g &\equiv& \frac{\langle U \rangle}{\langle K \rangle} = \frac{1}{6} \left(%
\frac{4}{3 \pi^2}\right)^{1/3} \frac{e^2 n_{0}^{1/3} \beta}{\varepsilon_0} 
\frac{\mathrm{Li}_{3/2}(-e^{\beta\mu_0})}{\mathrm{Li}_{5/2}(-e^{\beta\mu_0})}
\notag \\
&=& - \frac{\sqrt{\beta m_{e}/2}}{3^{4/3} \pi^{7/6}} \,\,\frac{e^2}{%
\varepsilon_0 \hbar} \frac{(\mathrm{Li}_{3/2}^{2}(-e^{\beta\mu_0}))^{2/3}}{%
\mathrm{Li}_{5/2}(-e^{\beta\mu_0})} \,,  \label{g}
\end{eqnarray}
covering both degenerate and non-degenerate systems, in the non-relativistic
regime. In the last equality in Eq. (\ref{g}) it was used the expression (%
\ref{muu}) of the equilibrium density in terms of the equilibrium fugacity $%
z = e^{\beta\mu_0}$ and the temperature $T$. In the dilute case, it follows
from the properties of the poly-logarithm function that $g \propto \langle U
\rangle/(\kappa_B T)$, while in the dense case $g \propto \langle U
\rangle/\varepsilon_F$, with $\mu_0 \approx \varepsilon_F$. 

\begin{figure}[!hbt]
\begin{center}
\includegraphics[width=8.0cm,height=6.0cm]{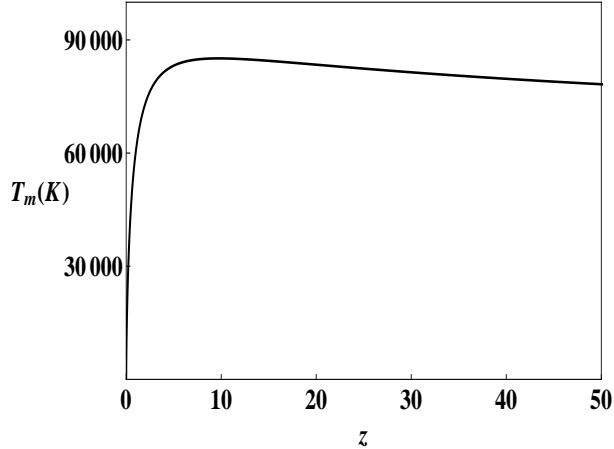}
\end{center}
\caption{Temperatures for the coupling parameter $g < 1$ in Eq. (\protect\ref%
{g}) are above the curve, as a function of the fugacity $z = e^{\protect\beta%
\protect\mu_0}$.}
\label{figure1}
\end{figure}

For both dilute or dense plasmas, the condition for low collisionality is
that the interaction energy should be small in comparison to the kinetic
energy, or $g \ll 1$ \cite{Ak}. Using Eq. (\ref{g}), the minimal temperature 
$T_m$ for low collisionality ($g < 1$, relaxing the inequality sign) for
both dilute and dense regimes follows from 
\begin{equation}
\kappa_B T > \kappa_B T_m \equiv \frac{m_e}{2 \times 3^{8/3} \pi^{7/3}}
\left(\frac{e^2}{\varepsilon_0 \hbar}\right)^2 \left(\frac{\mathrm{Li}%
_{3/2}^{4/3}(-e^{\beta\mu_0})}{\mathrm{Li}_{5/2}(-e^{\beta\mu_0})}\right)^2
\,.  \label{tmin}
\end{equation}
The result is shown in Fig. (\ref{figure1}), where $T > T_m$ is equivalent
to $g < 1$. Starting from $z \approx 0$ and increasing the density, larger
temperatures are needed for ideality, until reaching $z = 9.8, T = 8.5
\times 10^4 K$, corresponding to $n_0 = 2.9 \times 10^{29} m^{-3}$. For $z >
9.8$, smaller temperatures are admitted, due to the Pauli blocking effect
inhibiting collisions.

\begin{figure}[!hbt]
\begin{center}
\includegraphics[width=8.0cm,height=6.0cm]{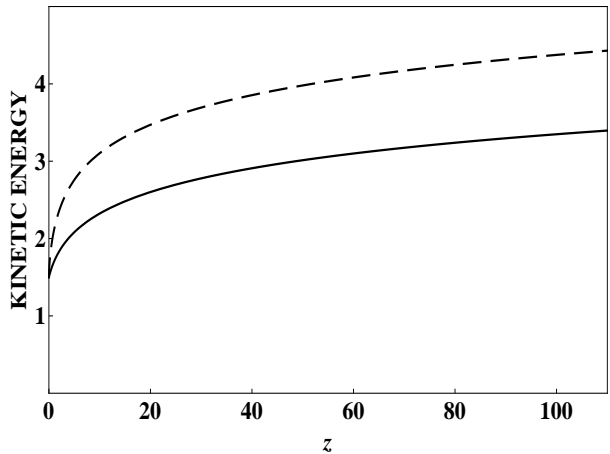}
\end{center}
\caption{Comparison between the average kinetic energy $\langle K \rangle$ -
the continuous curve - given by Eq. (\protect\ref{k}) and the simpler form $%
\langle K \rangle_{z}$ - the dashed curve - given by Eq. (\protect\ref{kz}),
as a function of the fugacity $z = e^{\protect\beta\protect\mu_0}$. Both
energies are normalized to $\protect\kappa_B T$.}
\label{figure2}
\end{figure}

For the sake of comparison, instead of 
\begin{equation}
\langle K\rangle =\frac{3}{2}\,\kappa _{B}T\,\frac{\mathrm{Li}%
_{5/2}(-e^{\beta \mu _{0}})}{\mathrm{Li}_{3/2}(-e^{\beta \mu _{0}})}\,,
\label{k}
\end{equation}%
Zamanian \textit{et al.} used \cite{Zamanian} the useful simpler expression 
\begin{equation}
\langle K\rangle _{Z}=\frac{3}{2}\kappa _{B}T+\frac{3}{5}\varepsilon _{F}
\label{kz}
\end{equation}%
as a measure of the kinetic energy per particle. More precisely, Ref. \cite%
{Zamanian} employed the arithmetic sum of the thermal and Fermi energies,
but in Eq. (\ref{kz}) we set some numerical factors to have agreement with
the exact form in the dilute and ultra-dense cases where one has resp. $%
\langle K\rangle \approx 3\,\kappa _{B}T/2$ and $\langle K\rangle \approx
3\,\varepsilon _{F}/5$. In fact, using Eq. (\ref{muu}) expressing the
density in terms of the fugacity and the temperature, as well as the
expression of the Fermi energy, one find 
\begin{equation*}
\frac{\langle K\rangle _{Z}}{\kappa _{B}T}=\frac{3}{2}+\frac{3^{5/3}}{10}%
\left( \frac{\pi }{2}\right) ^{1/3}\left[ -\mathrm{Li}_{3/2}(-e^{\beta \mu
_{0}})\right] ^{2/3},
\end{equation*}%
where the right-hand sides are functions of the fugacity only. This
expression is shown in Fig. (\ref{figure2}), compared to the more exact
result found from Eq. (\ref{k}). It is seen that the approximate form
overestimates the kinetic energy, due to slow convergence. Nevertheless, by
construction, for extreme degeneracy both quantities give the same numbers.

On the same spirit one can define a general electron thermal velocity (in
the sense of spreading of velocities) as $v_{T}\equiv \sqrt{2\langle
K\rangle /m_{e}}$, which is found from Eq. (\ref{k}), 
\begin{equation}
v_{T}=\left( \frac{3}{\beta m_{e}}\frac{\mathrm{Li}_{5/2}(-e^{\beta \mu
_{0}})}{\mathrm{Li}_{3/2}(-e^{\beta \mu _{0}})}\right) ^{1/2}\,.  \label{vt}
\end{equation}%
In the dilute case one has $v_{T}\approx \sqrt{3\kappa _{B}T/m_{e}}$, while
in the dense case $v_{T}\approx \sqrt{(6/5)\,\varepsilon _{F}/m_{e}}$.

For non-degenerate ions in strongly coupled plasma, the ion
crystallization effects \cite{r271,r272} that appear due to viscoelasticity
of the ion fluid in the ion momentum equation and cause damping of the
ion-acoustic wave are ignored under the assumptions (in three-dimensional version) $\tau
_{m} << \omega_{pi}$\ and $\partial {\bf u}_i/\partial t>> (\eta /\rho_{i})\nabla ^{2}{\bf u}_i$\ $+(1/\rho _{i})\left( \zeta +\eta /3\right) \nabla
\left( \mathbf{\nabla }\cdot{\bf u}_i\right) $, where $\rho _{i}=n_{i}m_{i}$%
\ is the ion mass density, $\tau _{m}$\ is the viscoelastic relaxation time
or  memory function for ions, $\eta $\ is the shear and $\zeta $\ are the
bulk ion viscosity coefficients, respectively.

\section{Linear Waves}

\subsection{Fluid theory}

We linearize the system given by equations (\ref{e1})-(\ref{e19}) by
considering the first order perturbations (with a subscript 1) relative to
the equilibrium, as follows, 
\begin{eqnarray}
n_{i} &=& n_{0} + n_{i1} \,, \quad n_{e} = n_{0} + n_{e1} \,, \quad u_{i} =
u_{i1} \,,  \notag \\
\phi &=& \phi _{1} \,, \quad \mu = \mu _{0}+\mu _{1} \,.
\end{eqnarray}

The dispersion relation is obtained assuming plane wave excitations $\sim 
\exp[i(k x - \omega t)]$, yielding 
\begin{equation}
\omega^{2} = \frac{\omega_{pi}^2 c_{s}^{2} k^{2} \left(1 + \frac{%
\alpha\,\hbar ^{2}k^{2}}{12\,m_{e}m_{i} c_{s}^2}\right)}{\omega_{pi}^2 +
\left(1 + \frac{\alpha\,\hbar ^{2}k^{2}}{12\,m_{e}m_{i} c_{s}^2}\right)
c_{s}^2 k^2}\,,  \label{e25}
\end{equation}
where 
\begin{equation}
c_{s} = \sqrt{\frac{1}{m_i}\left(\frac{\partial p}{\partial n_e}\right)_{0}}
= \sqrt{\frac{1}{\beta m_{i}}\frac{\mathrm{Li}_{3/2}(-e^{\beta \mu _{0}})}{%
\mathrm{Li}_{1/2}(-e^{\beta \mu _{0}})}}  \label{e22}
\end{equation}
plays the role of a generalized ion-acoustic speed and $\omega_{pi} = \sqrt{%
n_0 e^2/(m_i \varepsilon_0)}$ is the ion plasma frequency.

In the long wavelength limit $\alpha \,\hbar ^{2}k^{4}/(12m_{e}m_{i})\ll
c_{s}^{2}k^{2}\ll \omega _{pi}^{2}$ it follows from Eq.(\ref{e25}) that $%
\omega ^{2}\approx c_{s}^{2}k^{2}$. In the dilute case with a small fugacity 
$e^{\beta \mu _{0}}\ll 1$ the well-known classical result $c_{s}\approx
c_{sc}\equiv \sqrt{\kappa _{B}T/m_{i}}$ is verified. In the opposite,
extremely degenerate case where the fugacity $e^{\beta \mu _{0}}\gg 1$ one
find $c_{s}\approx \sqrt{(2/3)\varepsilon _{F}/m_{i}}$, which is the
ion-acoustic velocity for a three-dimensional ultra-dense plasma \cite{r19}.
Finally, the very short wavelength limit of the dispersion relation gives
ion oscillations such that $\omega =\omega _{pi}$, both in the classical or
quantum situations. This happens because the ions are no longer
shielded by electrons when wavelength is comparable to or smaller than the
electron shielding length. It is interesting to note that taking
the square root of both sides of Eq.(\ref{e25}) is identical to Eq.(4.5) in
Ref.\cite{r191} for the completely degenerate plasma case i.e., for $\alpha
=1/3$.

\begin{figure}[!hbt]
\begin{center}
\includegraphics[width=8.0cm,height=6.0cm]{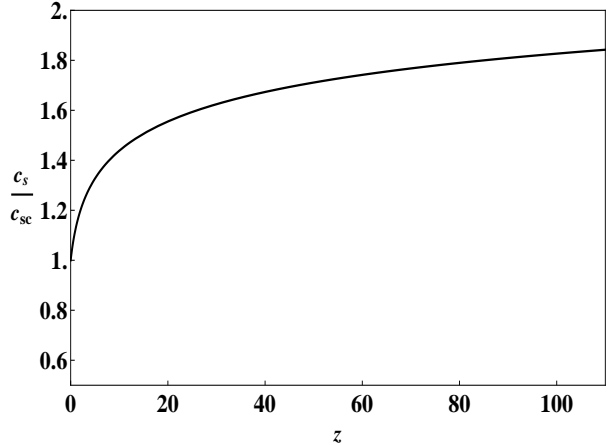}
\end{center}
\caption{The profile of the ion-acoustic speed $c_s$ from Eq. (\protect\ref%
{e22}), normalized to the classical expression $c_{sc}$, as a function of
the fugacity $z = e^{\protect\beta\protect\mu_0}$.}
\label{figure3}
\end{figure}

Using Eq. (\ref{e22}), the ion-acoustic speed $c_s$ normalized to the purely
classical expression $c_{sc}$ against $z = e^{\beta\mu_0}$ is shown
in Fig.(\ref{figure3}). It can be seen that as the value of $z$ increases
(i.e. the degeneracy of electrons and plasma density increase) the
ion-acoustic speed also increases. 

\subsection{Kinetic theory}

To endorse the macroscopic modeling, and to set the value of the parameter $%
\alpha$ in front of the quantum force, it is useful to compare with the
microscopic (quantum kinetic) results. Considering the Wigner-Poisson system 
\cite{r8} involving a cold ionic species and electrons, it is
straightforward to derive the linear dispersion relation 
\begin{equation}  \label{d1}
1 = \frac{\omega_{pi}^2}{\omega^2} + \frac{\omega_{pe}^2}{n_0} \int \frac{%
f_{0}(\mathbf{v}) \,d^3 v}{(\omega - \mathbf{k}\cdot\mathbf{v})^2 - \hbar^2
k^4/(4 m_{e}^2)} \,,
\end{equation}
where $f_{0}(\mathbf{v})$ is the equilibrium electronic Wigner function and $%
\omega_{pe} = \sqrt{n_0 e^2/(m_e \varepsilon_0)}$ is the electron plasma
frequency.

The longitudinal response of an electron-ion plasma in a Fermi-Dirac
equilibrium 
\begin{equation}
f_{0}(\mathbf{v}) = \frac{A}{1 + e^{\beta(\varepsilon - \mu_0)}}
\end{equation}
has been calculated in \cite{mm}, where $A$ and $\mu_0$ are obtained from
Eq. (\ref{muu}). Including the first order correction from quantum recoil,
the result is 
\begin{eqnarray}
1 = \frac{\omega_{pi}^2}{\omega^2} &-& \frac{\omega_{pi}^2}{c_{s}^2 k^2}%
\Bigl[1  \label{m} \\
&-& \frac{m_e \left(\omega^2 + \hbar^2 k^4/(12\, m_{e}^2)\right)}{k^2
\kappa_B T}\,\frac{\mathrm{Li}_{-1/2}(-e^{\beta\mu _{0}})}{\mathrm{Li}%
_{1/2}(-e^{\beta\mu _{0}})}\Bigr] \,,  \notag
\end{eqnarray}
which follows from Eq. (29) of \cite{mm}, in a different notation. The first
and second terms in the right-hand side of Eq.(\ref{m}) are, respectively,
the ionic and electronic res\-pon\-ses of the plasma.

For the treatment of low-frequency waves, for sim\-pli\-ci\-ty it is
sufficient to consider the static electronic response, so that we set $%
\omega \approx 0$ in the last term of Eq. (\ref{m}). From inspection, and
since we want to retain the first order quantum correction, this
approximation requires $\omega^2 \ll \hbar^2 k^4/(12\, m_{e}^2)$. Under the
long wavelength assumption $k^2 c_{s}^2 \ll \omega_{pi}^2$ and the leading
order result $\omega^2 \approx c_{s}^2 k^2$, it follows that 
\begin{equation}  \label{range}
\frac{12\, m_{e}^2 c_{s}^2}{\hbar^2} \ll k^2 \ll \frac{\omega_{pi}^2}{c_{s}^2%
} \,.
\end{equation}
Taking into account the ion-acoustic velocity from Eq. (\ref{e22}), from Eq.
(\ref{range}) one has the necessary condition 
\begin{equation}
\frac{\beta^2 \hbar^2 \omega_{pi}^2}{12} \gg \left(\frac{m_{e}}{m_i}%
\right)^2 \left(\frac{\mathrm{Li}_{3/2}(-e^{\beta\mu _{0}})}{\mathrm{Li}%
_{1/2}(-e^{\beta\mu _{0}})}\right)^2 \,.  \label{lf}
\end{equation}
The combined low-frequency and long wavelength requirement (\ref{lf}) is
more easily worked out in the dilute ($e^{\beta\mu_0} \ll 1$) and fully
degenerate ($e^{\beta\mu_0} \gg 1$) cases. For hydrogen plasma and using the
appropriate asymptotic expansions of the poly-logarithm functions, one find $%
n_0/T^2 \gg 3.5 \times 10^{16} \,m^{-3} K^{-2}$ in the non-degenerate
si\-tua\-tion, and $n_0 \ll 4.5 \times 10^{37} \,m^{-3}$ for very dense
systems. It is seen that non-degenerate plasmas satisfy (\ref{lf}) more
easily in denser and colder plasmas, while fully degenerate plasmas safely
fit the assumptions, except for extreme densities (e.g neutron star), which
would deserve a relativistic treatment. Otherwise, there would be the need
to retain the full electronic response in Eq.(\ref{m}). As a consequence, a
somewhat more involved dispersion relation would be found. In fact, using $%
n_0$ from Eq. (\ref{muu}), it can be shown that the necessary condition (\ref%
{lf}) is safely attended for all fugacities, as far as $T \ll 10^9 \,K$,
which is reasonable in view of the non-relativistic assumption.

Dropping $\omega$ in the electronic response, Eq. (\ref{m}) considerably
simplify, reducing to 
\begin{equation}  \label{sr}
1 = \frac{\omega_{pi}^2}{\omega^2} - \frac{\omega_{pi}^2}{c_{s}^2 k^2}%
\left(1 - \frac{\hbar^2 k^2}{12\, m_{e} \kappa_B T}\,\frac{\mathrm{Li}%
_{-1/2}(-e^{\beta\mu _{0}})}{\mathrm{Li}_{1/2}(-e^{\beta\mu _{0}})}\right)
\,.
\end{equation}
Solving for the frequency yield 
\begin{equation}
\omega^2 = \frac{\omega_{pi}^2 c_{s}^2 k^2}{\omega_{pi}^2 + \left(1 - \frac{%
\beta^2 \hbar^2 \omega_{pe}^2}{12}\frac{\mathrm{Li}_{-1/2}(-e^{\beta\mu_{0}})%
}{\mathrm{Li}_{3/2}(-e^{\beta\mu_{0}})}\right) c_{s}^2 k^2} \,.  \label{ok}
\end{equation}

The expression from kinetic theory is valid for wavelengths larger
than the electron shielding length of the system. To make a comparison with
the result from fluid theory, it is necessary to expand (\ref{ok}) for small
wavenumbers, 
\begin{eqnarray}
\omega ^{2} &=&c_{s}^{2}k^{2}\!\left[ 1+\left( -1+\frac{\beta ^{2}\hbar
^{2}\omega _{pe}^{2}}{12}\frac{\mathrm{Li}_{-1/2}(-e^{\beta \mu _{0}})}{%
\mathrm{Li}_{3/2}(-e^{\beta \mu _{0}})}\right) \frac{c_{s}^{2}k^{2}}{\omega
_{pi}^{2}}\right]   \notag \\
&+&\mathcal{O}(k^{6})\,.  \label{kkk}
\end{eqnarray}

Next, expand the fluid theory expression (\ref{e25}) for small wavenumbers, 
\begin{equation}  \label{fff}
\omega^2 = c_{s}^2 k^2 \left[1 + \left(- 1 + \frac{\alpha \hbar^2
\omega_{pi}^2}{12 m_e m_i c_{s}^4}\right)\frac{c_{s}^2 k^2}{\omega_{pi}^2}%
\right] + \mathcal{O}(k^6) \,.
\end{equation}

Equations (\ref{kkk}) and (\ref{fff}) are equivalent provided we set 
\begin{equation}  \label{alp}
\alpha = \frac{\mathrm{Li}_{3/2}(-e^{\beta\mu_{0}})\,\mathrm{Li}%
_{-1/2}(-e^{\beta\mu_{0}})}{[\mathrm{Li}_{1/2}(-e^{\beta\mu_{0}})]^2} \,,
\end{equation}
which is our ultimate choice. Therefore, to comply with the results of
kinetic theory on quantum ion-acoustic waves in a three-dimensional
Fermi-Dirac equilibrium, the numerical coefficient in the quantum force has
to be a function of the fugacity. In particular, with $z = \exp(\beta\mu_0)$%
, we have $\alpha \rightarrow 1$ for $z \rightarrow 0$ and $\alpha
\rightarrow 1/3$ as $z \rightarrow \infty$. Moreover, as seen in Fig. (\ref%
{figure4}), the coefficient $\alpha$ is a monotonically decreasing function
of the fugacity, showing that the quantum force becomes less effective in
denser systems. The result $\alpha \to 1$ for non-degenerate systems agrees
with the quantum hydrodynamic model for semiconductor devices derived in 
\cite{Gardner}, while $\alpha \to 1/3$ agrees with \cite{bon, akba} in the
fully degenerate case. On the other hand, high frequency waves such as
quantum Langmuir waves would be correctly described by a value $\alpha = 3$,
in order to reproduce the Bohm-Pines \cite{bp} dispersion relation $\omega^2
= \omega_{pe}^2 + (3/5)\,k^2 v_{F}^2 + (1/4) \hbar^2 k^4/m_{e}^2$, where $%
v_F = (2 E_F/m_e)^{1/2}$ is the Fermi velocity. 
\begin{figure}[!hbt]
\begin{center}
\includegraphics[width=8.0cm,height=6.0cm]{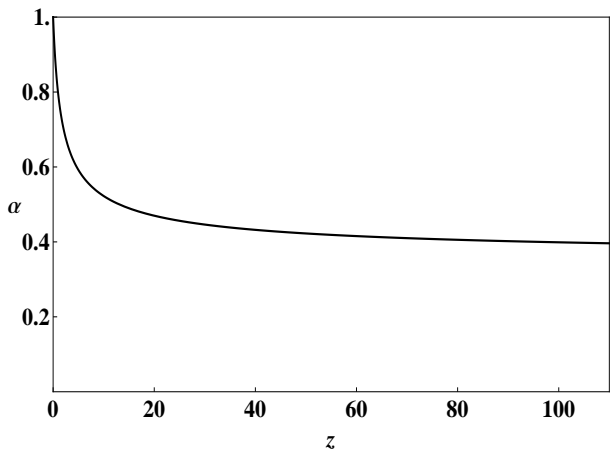}
\end{center}
\caption{Behavior of the numerical coefficient $\protect\alpha$ in Eq. (%
\protect\ref{alp}), as a function of the fugacity $z = \exp(\protect\beta%
\protect\mu_0)$.}
\label{figure4}
\end{figure}

The detailed account of the collisionless damping of quantum ion-acoustic
waves has been considered in \cite{mm}, where the damping rate is shown to
be small, as long as the ion temperature is much smaller than the electron
temperature of the plasma.

\section{Nonlinear waves}

Having performed the analysis of linear quantum ion-acoustic waves, it is
worth to consider the nonlinear structures which are accessible through our
hydrodynamic model.

From now on, it is useful to define the rescaling 
\begin{eqnarray}
\tilde{x} &=& \frac{\omega_{pi}x}{c_{s}} \,, \quad \tilde{t} = \omega_{pi} t
\,, \quad \tilde{n}_{e,i} = \frac{n_{e,i}}{n_0} \,,  \notag \\
\tilde{u}_{i} &=& \frac{u_i}{c_{s}} \,, \quad \Phi = \frac{e\phi}{%
m_{i}c_{s}^2} \,,
\end{eqnarray}
so that the model equations (\ref{e1})-(\ref{e4}) can be written in a
normalized form as follows, 
\begin{eqnarray}
\frac{\partial\tilde{n}_{i}}{\partial \tilde{t}} + \frac{\partial }{\partial 
\tilde{x}}(\tilde{n}_{i}\tilde{u}_{i}) &=& 0 \,,  \label{e30} \\
\frac{\partial \tilde{u}_{i}}{\partial \tilde{t}} + \tilde{u}_{i} \frac{%
\partial}{\partial\tilde{x}}\tilde{u}_{i} &=& - \frac{\partial\Phi}{\partial 
\tilde{x}} \,,  \label{e31} \\
0 = \frac{\partial \Phi }{\partial \tilde{x}} &-& \frac{\mathrm{Li}%
_{1/2}(-e^{\beta \mu _{0}})}{\mathrm{Li}_{1/2}(-e^{\beta \mu })}\frac{%
\partial \tilde{n}_{e}}{\partial \tilde{x}}  \notag \\
&+& \frac{H^{2}}{2}\frac{\partial}{\partial \tilde{x}}\left( \frac{1}{\sqrt{%
\tilde{n}_{e}}}\frac{\partial^{2}}{\partial \tilde{x}^{2}} \sqrt{\tilde{n}%
_{e}}\right) \,,  \label{e32} \\
\frac{\partial^{2}\Phi }{\partial\tilde{x}^{2}} &=& \tilde{n}_{e} - \tilde{n}
_{i} \,,  \label{e33}
\end{eqnarray}
introducing the quantum parameter $H$ given by 
\begin{equation}  \label{hhh}
H = \frac{\beta\hbar\omega_{pe}}{\sqrt{3}} \left(\frac{\mathrm{Li}%
_{-1/2}(-e^{\beta\mu_0})}{\mathrm{Li}_{3/2}(-e^{\beta\mu_0})}\right)^{1/2}
\,.
\end{equation}
In the dilute or fully degenerate cases one has resp. $H \approx
\beta\hbar\omega_{pe}/\sqrt{3}$ or $H \approx \hbar\omega_{pe}/(2
\varepsilon_F)$. Moreover, from Eq.(\ref{e18}), 
\begin{equation}
\tilde{n}_{e} = \frac{\mathrm{Li}_{3/2}(-e^{\beta\mu})}{\mathrm{Li}%
_{3/2}(-e^{\beta\mu_{0}})} \,.  \label{e34}
\end{equation}%
In the following, for simplicity the tilde will be omitted in the normalized
quantities.

In order to find a nonlinear evolution equation, a stretching of the
independent variables $x,$ $t$ is defined as follows \cite{r12,r18},%
\begin{equation*}
\xi =\varepsilon ^{1/2}(x-V_{0}t)\,,\quad \tau =\varepsilon ^{3/2}t,
\end{equation*}%
where $\varepsilon $ is a small parameter and $V_{0}$ is the phase velocity
of the wave to be determined later on. The perturbed quantities can be
expanded in powers of $\varepsilon $, 
\begin{eqnarray}
n_{i} &=& 1+\varepsilon n_{i1}+\varepsilon ^{2}n_{i2} + \dots \,,  \notag \\
n_{e} &=& 1+\varepsilon n_{e1}+\varepsilon ^{2}n_{e2} + \dots \,,  \notag \\
u_{i} &=& \varepsilon u_{i1}+\varepsilon ^{2}u_{i2} + \dots \,,  \notag \\
\Phi &=& \varepsilon \Phi _{1}+\varepsilon ^{2}\Phi _{2} + \dots \,,  \notag
\\
\mu &=& \mu _{0}+\varepsilon \mu _{1}+\varepsilon ^{2}\mu _{2} + \dots
\label{e37}
\end{eqnarray}

The lowest order equations give 
\begin{equation}
n_{i1}=u_{i1}=n_{e1}=\Phi _{1} \,,  \label{e38}
\end{equation}
and 
\begin{equation}
V_{0}=\pm 1 \,,  \label{e39}
\end{equation}
which is the normalized phase velocity of the ion-acoustic wave in plasmas
with arbitrary degeneracy of electrons. From now on, we set $V_{0}=1$
without loss of generality.

Now collecting the next higher order terms, we have 
\begin{eqnarray}
\frac{\partial n_{i2}}{\partial \xi } &=& \frac{\partial n_{i1}}{\partial
\tau }+ \frac{\partial }{\partial \xi }(n_{i1}u_{i1}) +  \notag \\
&+&\frac{\partial u_{i1}}{\partial \tau }+u_{i1}\frac{\partial u_{i1}}{%
\partial \xi }+\frac{\partial \Phi _{2}}{\partial \xi } \,,  \label{e40} \\
\frac{\partial n_{e2}}{\partial \xi } &=& \frac{\partial \Phi _{2}}{\partial
\xi }+ \alpha n_{e1}\frac{\partial n_{e1}}{\partial \xi} + \frac{H^{2}}{4}%
\frac{\partial ^{3}n_{e1}}{\partial \xi ^{3}} \,.  \label{e41}
\end{eqnarray}
Using the next higher order Poisson's equation together with Eqs. (\ref{e38}%
), (\ref{e40}) and (\ref{e41}) yield the KdV equation for ion-acoustic waves
in plasmas with arbitrary degeneracy of electrons,%
\begin{equation}
\frac{\partial \Phi _{1}}{\partial \tau }+a\Phi _{1}\frac{\partial \Phi _{1}%
}{\partial \xi }+b\frac{\partial ^{3}\Phi _{1}}{\partial \xi ^{3}}=0,
\label{e42}
\end{equation}%
where the nonlinear and dispersive coefficients $a$ and $b$ are resp.
defined as 
\begin{equation}
a = \frac{3 - \alpha}{2} \,, \quad b = \frac{1}{2}\left(1-\frac{H^{2}}{4}%
\right) \,.  \label{e44}
\end{equation}
In the fully classical limit (non-degeneracy and no Bohm potential), one has 
$a = 1, b = 1/2$, recovering the KdV equation for classical ion-acoustic
waves \cite{r28}. The effect of arbitrary degeneracy of electrons appears in
both the nonlinear and dispersive coefficients in the KdV equation (\ref{e42}%
).

It is easy to derive traveling wave solutions for the problem. One of them
is the one-soliton solution of the KdV equation (\ref{e42}) given by 
\begin{equation}
\Phi _{1}=D\,\mathrm{sech}^{2}(\frac{\eta }{W})\,,  \label{e45}
\end{equation}%
where $D=3u_{0}/a$ and $W=\sqrt{4b/u_{0}}$ are resp. the height and width of
the soliton. The polarity of the soliton depends on the sign of $D$. In the
co-moving frame one has $\eta =\xi -u_{0}\tau $, where $u_{0}$ is the speed
of the nonlinear structure. Decaying boundary conditions in the co-moving
system were used. For a given perturbation speed, one conclude that larger
degeneracy (larger $a,b$) gives a smaller scaled amplitude and a
larger scaled width. This is because it becomes harder to
accommodate more fermions in a localized wave packet under strong
degeneracy. The transformed coordinate $\eta $\ can be
written as $\eta =\varepsilon ^{1/2}\tilde{\eta} $ where $\tilde{%
\eta}=x-Vt$\textbf{\ and }$V=V_{0}+\varepsilon u_{0}$ is the
soliton velocity in the lab frame.

It can be seen from the relation (\ref{e44}) that the dispersive coefficient 
$b$ disappears at $H=2$. In principle, the lack of a dispersive term
eventually yields the formation of a shock. However, actually in this case a
dispersive contribution could be obtained from a higher-order perturbation
theory, as occurs in the Kawahara equation \cite{Kawahara}. In the present
context of quantum ion-acoustic nonlinear waves, the soliton solution can
exist only for $H\neq 2$, with a proper balance between dispersion and
nonlinearity. Notice that for $H<2$ the soliton velocity is positive i.e., $%
u_{0}>0$ (which means $V>V_{0}$ and it moves with
supersonic speed) and we have a hump (bright) soliton structure since $a>0$
and $D>0$. However, for $H>2$ case the dispersive coefficient becomes
negative i.e., $b<0$, so that the soliton solution will exist only if $%
u_{0}<0$ (i.e., $V<V_{0}$ soliton moves with subsonic speed), since
the width $W$ should have real values. As $u_{0}$ is negative in the $H>2$
case, the nonlinearity coefficient remains positive i.e., $a>0$, therefore $%
D<0$ which gives a dip (or dark) soliton instead of a hump (or bright)
structure \cite{r29}. In brief, the model predicts hump solitons for $H<2$
case and dip solitons for $H>2$. Finally, in the special fine tuning case
with $H=2$ there is a shock instead of solitonic solutions, at least within
the present order of perturbation theory.

\section{On bright and dark propagating solitons}

\begin{figure}[!hbt]
\begin{center}
\includegraphics[width=8.0cm,height=6.0cm]{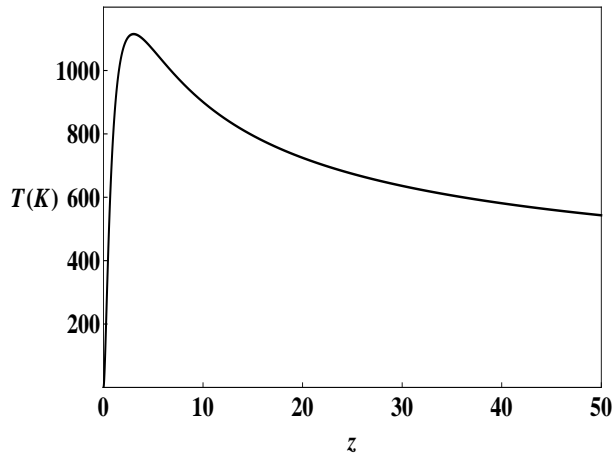}
\end{center}
\caption{The temperatures for $H > 2$ satisfying Eq. (\protect\ref{max}) are
below the curve, where $z = e^{\protect\beta\protect\mu_0}$.}
\label{figure5}
\end{figure}

The qualitative differences of quantum ion-acoustic soliton propagation for $%
H<2$ or $H>2$ deserve a closer examination about the associated physical
conditions. First, the quantum parameter in Eq. (\ref{hhh}) can be
re-expressed according to 
\begin{equation}
H^{2}=-\frac{1}{3\pi }\left( \frac{m_{e}}{2\pi \kappa _{B}T}\right) ^{1/2}%
\frac{e^{2}}{\varepsilon _{0}\hbar }\,\,\mathrm{Li}_{-1/2}(-e^{\beta \mu
_{0}})\,,  \label{newh}
\end{equation}%
where the equilibrium density in Eq. (\ref{muu}) was employed. From the last
equation, one find that $H>2$ occurs for sufficiently small temperatures, or 
\begin{equation}
\kappa _{B}T<\frac{m_{e}}{288\pi ^{3}}\left( \frac{e^{2}}{\varepsilon
_{0}\hbar }\right) ^{2}\,\,\left[ \mathrm{Li}_{-1/2}(-e^{\beta \mu _{0}})%
\right] ^{2}\,,  \label{max}
\end{equation}%
as illustrated in Fig. (\ref{figure5}). Low temperature plasmas with $%
T<10^{3}K$ are therefore candidates for the peculiar dark solitons. Starting
from $z=e^{\beta \mu _{0}}\approx 0$ the maximal temperature increases until 
$z=3,T=1.1\times 10^{3}K$ (corresponding to $n_{0}=3.0\times 10^{26}m^{-3}$%
), when it starts to decrease.

\begin{figure}[!hbt]
\begin{center}
\includegraphics[width=8.0cm,height=6.0cm]{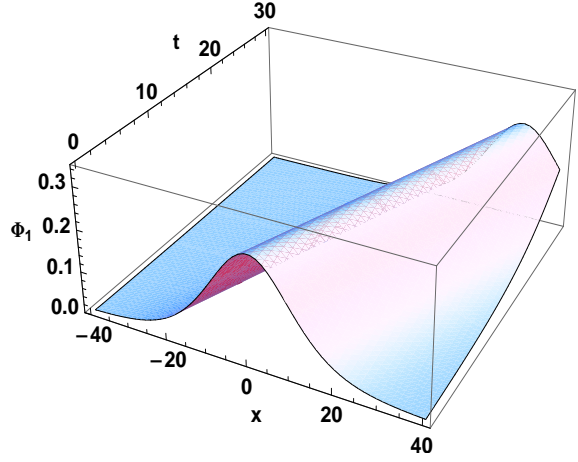}
\end{center}
\caption{Hump soliton structure for $H<2$ is shown moving with supersonic
speed in lab frame. The soliton hump corresponds to $T=10^{5}K$, $z=5$, $%
u_{0}=0.1$, $\protect\varepsilon =0.1$ 
for which $H=0.64$, $%
n_{0}=3.5\times 10^{29}m^{-3}$, $\protect\omega _{pe}=3.3\times
10^{16}s^{-1}$, respectively.}
\label{figure6}
\end{figure}

\begin{figure}[!hbt]
\begin{center}
\includegraphics[width=8.0cm,height=6.0cm]{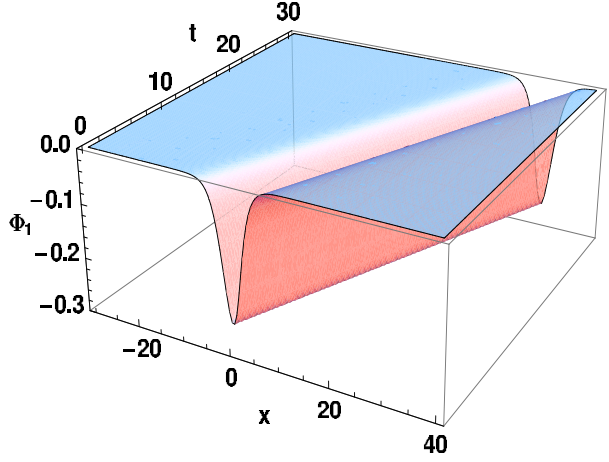}
\end{center}
\caption{Dip soliton structure for $H > 2$ is shown moving  with
subsonic speed in lab frame. The soliton dip is obtained at $T=10^{3}K$, $%
z=5 $, $u_{0} = - 0.1$, 
for which $H=2.03$, $n_{0}
= 3.5 \times 10^{26}m^{-3}$, $\protect\omega _{pe} = 1.1\times 10^{15}s^{-1}$%
, respectively.}
\label{figgure7}
\end{figure}

As an example, in Figs.(\ref{figure6}) and(\ref{figgure7}) the two
classes of quantum ion-acoustic bright or dark solitons are shown, following
Eq. (\ref{e45}). The bright soliton ($H<2$) moves with supersonic
speed while the dark soliton ($H>2$) moves with subsonic speed. 

\begin{figure}[tbp]
\par
\begin{center}
\includegraphics[width=8.0cm,height=6.0cm]{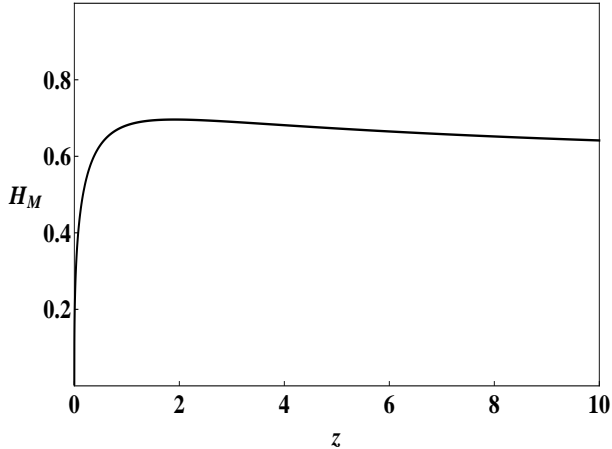}
\end{center}
\caption{Maximal quantum parameter $H_M$ satisfying the weak coupling
assumption, according to Eq. (\protect\ref{hmax}), where $z = e^{\protect%
\beta\protect\mu_0}$.}
\label{figure8}
\end{figure}

On the other hand, it is interesting to examine the conditions for weak coupling 
as deduced in the present theory.
Combining the weak coupling condition yielding
the minimal temperature in Eq. (\ref{tmin}) with Eq. (\ref{newh}) gives an
upper bound on the quantum diffraction parameter, or 
\begin{equation}
H^{2}<H_{M}^{2}\equiv -\left( \frac{3}{\pi }\right) ^{1/3}\frac{\mathrm{Li}%
_{-1/2}(-e^{\beta \mu _{0}})\mathrm{Li}_{5/2}(-e^{\beta \mu _{0}})}{[\mathrm{%
Li}_{3/2}^{2}(-e^{\beta \mu _{0}})]^{2/3}}\,,  \label{hmax}
\end{equation}%
which is shown in Fig. (\ref{figure8}). It follows that large $H>2$%
\ values fall within the strongly coupled regime where coupling
parameter $g$ for degenerate electrons may become near or greater
than one.

Nevertheless, considering ion-acoustic waves at least, a strong coupling between electrons will be not the main aspect of the dynamics. Although the complete analysis of the strongly coupled plasma regime is beyond the scope of this work, some conclusions can be found from the simplest way to introduce non-ideality for electrons, namely the addition of a dissipation term $-\omega_{coll} u_e$ in the right-hand side of Eq. (\ref{e3}), where $u_e$ is the electron fluid velocity and $\omega_{coll}$ is the electron-electron collision frequency. Using the continuity equation for electrons it is possible to estimate $\omega n_{e1} \approx k\, n_0 u_{e1}$, where $n_{e1}$ and $u_{e1}$ are the first-order perturbations of the electron fluid density and velocity. Finally with $\omega \approx c_s k$ one finds that the dissipation term is negligible with respect to the pressure term provided $\omega \gg (m_e/m_i) \omega_{coll}$, which is always satisfied within the inertialess electrons assumption. We note that according to the Landau expression \cite{r30} one has in the non-degenerate case 
\begin{equation}
\label{cf}
\frac{\omega_{coll}}{\omega_{pe}} \approx \frac{\ln\Lambda}{\Lambda} \,,
\end{equation}
where $\Lambda \sim 1/g^{3/2}$ is the plasma parameter. In the fully degenerate case the right-hand side of Eq. (\ref{cf}) needs to be multiplied by the Pauli blocking factor $\kappa_B T/\varepsilon_F$. The conclusion is that except for very high $g \gg 1$ the electron-electron coupling can be neglected as long as the inertialess assumption is valid.

\section{Conclusion}

The linear and nonlinear ion-acoustic waves in a non-relativistic quantum
plasma with arbitrary degeneracy of electrons have been investigated.
Besides degeneracy, the quantum diffraction effect of electrons was also
included, in terms of the Bohm potential. The linear dispersion relation for
quantum ion-acoustic waves was found in terms of a generalized ion-acoustic
speed, valid for both the dilute and dense cases. The numerical factor $%
\alpha $ in front of the quantum force in the macroscopic model was fixed in
order to comply with the kinetic theory results. The corresponding KdV
equation was obtained using the reductive perturbation method. The possible
classes of propagating solitons, namely bright for $H<2$ %
moving with supersonic speed and dark for $H>2$ case
moving with subsonic speed were discussed, where $H$ is a measure of the
strength of quantum diffraction effects arising from the Bohm potential.
To conclude, the derivation covers both the basic
quantum effects in plasmas (arising resp. from quantum statistics and
wave-like behavior of the charge carriers), in both the dilute and dense
regimes. For instance, from Eq. (\ref{e45}) the scaled amplitude of
the soliton becomes smaller for larger degeneracy, with $D=3u_{0}$ for $%
\alpha =1$ (non-degenerate case) and $D=9u_{0}/4$ for $\alpha =1/3$ (fully
degenerate) case. The results are useful for the understanding of
ion-acoustic wave propagation in an unmagnetized quantum plasma with
arbitrary degeneracy of electrons.

\acknowledgments

FH acknowledges CNPq (Conselho Nacional de Desenvolvimento Cient\'{\i}fico e
Tecnol\'ogico) for financial support. SM acknowledges CNPq and TWAS (The
World Academy of Sciences) for a CNPq-TWAS postdoctoral fellowship.


\begin{thebibliography}{99}
\bibitem{r2} S. H. Glenzer, O. L. Landen, P. Neumayer and R. W. Lee, \textit{%
Phys. Rev. Lett.} \textbf{98}, 065002 (2007).



\bibitem{r5} A. K. Harding and D. Lai, \textit{Rep. Prog. Phys.} \textbf{69}%
, 2631 (2006).

\bibitem{r1} A. J\"ungel, \textit{Trans\-port E\-qua\-tions for
Se\-mi\-con\-ductors} (Springer, Berlin, 2009).




\bibitem{r8} F. Haas, \textit{Quantum Plasmas: an Hydrodynamic Approach}
(Springer, New York, 2011).

\bibitem{mel} D. Melrose, \textit{Quantum Plasmadynamics - Unmagnetized
Plasmas} (Springer-Verlag, New York, 2008).



\bibitem{r12} F. Haas, L. G. Garcia, J. Goedert and G. Manfredi, \textit{%
Phys. Plasmas} \textbf{10}, 3858 (2003).

\bibitem{r13} G. Manfredi and F. Haas, \textit{Phys. Rev. B} \textbf{64},
075316 (2001).

\bibitem{r14} P. K. Shukla and S. Ali, \textit{Phys. Plasmas} \textbf{12},
114502 (2005).

\bibitem{r15} R. Sabry, W. M. Moslem and P. K. Shukla, \textit{Phys. Lett. A}
\textbf{372}, 5691 (2008).

\bibitem{r16} U. M. Abdelsalam, W. M. Moslem and P. K. Shukla, \textit{Phys.
Plasmas} \textbf{15}, 052303 (2008).

\bibitem{r17} A. E. Dubinov and A. A. Dubinova, \textit{Plasma Phys. Rep.} 
\textbf{33}, 859 (2007).

\bibitem{r18} S. Mahmood and F. Haas, \textit{Phys. Plasmas} \textbf{21},
102308 (2014).

\bibitem{r11} A. E. Dubinov and A. A. Dubinova, \textit{Plasma Phys. Rep.} 
\textbf{34}, 403 (2008).

\bibitem{solid} G. Manfredi and J. Hurst, \textit{Plasma Phys. Control.
Fusion} \textbf{57}, 054004 (2015).


\bibitem{gregori} J. E. Cross, B. Reville and G. Gregori, \textit{Astrophys.
J.} \textbf{795}, 59 (2014).

\bibitem{r19} N. Maafa, Phys. Scripta \textbf{48}, 351 (1993).




\bibitem{r191} B. Eliasson and P.K. Shukla, J. Plasma Phys. 76, \textbf{7}
(2010).

\bibitem{r22} A. Mushtaq and D. B. Melrose, \textit{Phys. Plasmas} \textbf{16%
}, 102110 (2009).

\bibitem{mm} D. B. Melrose and A. Mushtaq, \textit{Phys, Rev. E} \textbf{82}%
, 056402 (2010).

\bibitem{r23} B. Eliasson and P. K. Shukla, \textit{Phys. Scripta} \textbf{78%
}, 025503 (2008).

\bibitem{r24} A. E. Dubinov, A. A. Dubinova and M. A. Sazokin, \textit{J.
Commun. Tech. Elec.} \textbf{55}, 907 (2010).


\bibitem{Barker} J. R. Barker and D. K. Ferry, \textit{Semicond. Sci.
Technol.} \textbf{13}, A135 (1998).

\bibitem{Gardner} C. L. Gardner, \textit{SIAM J. Appl. Math.} \textbf{54},
409 (1994).

\bibitem{Grubin} H. L. Rubin, T. R. Govindan, J. P. Kreskovski and M. A.
Stroscio, \textit{Solid St. Electron.} \textbf{36}, 1697 (1993).


\bibitem{r26} R. K. Pathria and P. D. Beale, \textit{Statistical Mechanics -
3rd ed.} (Elsevier, New York, 2011).

\bibitem{r27} L. Lewin, \textit{Poly\-lo\-ga\-rithms and Asso\-cia\-ted
Functions} (North Holland, New York, 1981).

\bibitem{Ak} A. I. Akhiezer, I. A. Akhiezer, R. V. Polovin, A. G. Sitenko
and K. N. Stepanov, \textit{Plasma Electrodynamics - vol. I} (Pergamon,
Oxford, 1975).

\bibitem{Zamanian} J. Zamanian, M. Marklund and G. Brodin, \textit{New J.
Phys} \textbf{12}, 043019 (2010).

\bibitem{r271} P. K. Shukla and B. Eliasson, \textit{Rev. Mod. Phys}. 
\textbf{83}, 885(2011).

\bibitem{r272} A. P. Misra and P. K. Shukla, \textit{Phys. Rev. E} \textbf{85%
}, 026409 (2012).

\bibitem{bon} D. Michta, F. Graziani and M. Bonitz, \textit{Contrib. Plasma
Phys.} \textbf{55}, 437 (2015).

\bibitem{akba} M. Akbari-Moghanjoughi, \textit{Phys. Plasmas} \textbf{22},
022103 (2015).

\bibitem{bp} D. Bohm and D. Pines, \textit{Phys. Rev.} \textbf{92}, 609
(1953).

\bibitem{r28} R. C. Davidson, \textit{Methods in Nonlinear Plasma Theory}
(Academic Press, New York, 1972).

\bibitem{Kawahara} T. Kawahara, \textit{Phys. Soc. Japan} \textbf{33}, 260
(1972).

\bibitem{r29} V. Yu. Belashov and S. V. Vladimirov, \textit{Solitary Waves
in Dispersive Complex Media} (Springer-Verlag, Berlin-Heidelberg, 2005).

\bibitem{r30} E. M. Lifshitz and L. P. Pitaevskii, \textit{Physical Kinetics} (Pergamon, Oxford, 1981).
\end{thebibliography}
\end{document}